\documentclass[12pt,linenumbers, preprint]{aastex}

\usepackage{amsmath,longtable,natbib}
\usepackage{float}
\usepackage{hyperref}
\usepackage{adjustbox}
\usepackage{caption}
\usepackage{graphicx,epstopdf}
\usepackage[usenames,dvipsnames]{color}

\DeclareGraphicsExtensions{.ps}
\shorttitle{}
\shortauthors{Sharma et al.}
\usepackage{subfig}
\begin{document}
\captionsetup[figure]{labelfont={bf},labelformat={simple},labelsep=period,name={Figure}}
\captionsetup[table]{labelfont={bf},labelformat={simple},labelsep=period,name={Table}}
\title{Retracing the Cold Plasma Dispersion Law in Pulsar B0329+54: New Insights into Frequency-Dependent Dispersion Measures}
\author{
Shyam~S.~Sharma\altaffilmark{1,2}, Tetsuya Hashimoto\altaffilmark{2}, Tomotsugu Goto\altaffilmark{3}, Shotaro Yamasaki\altaffilmark{2}, Simon C.-C. Ho\altaffilmark{4,5,6,7}}
\altaffiltext{1}{Institute of Astronomy and Astrophysics, Academia Sinica, Taipei 10617, Taiwan}
\altaffiltext{2}{Department of Physics, National Chung Hsing University, Taichung 40227, Taiwan}
\altaffiltext{3}{Institute of Astronomy, National Tsing Hua University, Hsinchu 30013, Taiwan}
\altaffiltext{4}{Research School of Astronomy and Astrophysics, The Australian National University, Canberra, ACT 2611, Australia}
\altaffiltext{5}{Centre for Astrophysics and Supercomputing, Swinburne University of Technology, Hawthorn, VIC 3122, Australia}
\altaffiltext{6}{OzGrav: The Australian Research Council Centre of Excellence for Gravitational Wave Discovery, Hawthorn, VIC 3122, Australia}
\altaffiltext{7}{ASTRO3D: The Australian Research Council Centre of Excellence for All-sky Astrophysics in 3D, ACT 2611, Australia}

\section*{ABSTRACT}
\vspace{-3.9mm}

{Multiple studies have investigated potential frequency-dependent dispersion measures (DM) in PSR B0329+54, with sensitivities at levels of 10$^{-3}$ pc cm$^{-3}$ or higher, using frequencies below 1 GHz. Utilizing the extensive bandwidth of the upgraded Giant Meterwave Radio Telescope, we conducted simultaneous observations of this pulsar across a frequency range of 300 to 1460 MHz. Our observations reveal a distinct point in the pulse profile of PSR B0329+54 that appear to aligns remarkably well with the cold-plasma dispersion law, resulting in a unique measured DM across the entire frequency range. In contrast, using times of arrival (ToAs) from widely adopted pulsar timing techniques (e.g., FFTFIT)—leads to frequency-dependent DMs. We investigated the potential causes of these frequency-dependent DMs in this pulsar and their relationship with the underlying magnetic field geometry corresponding to the radio emission. Our study reveals that all frequencies in the range 300–1460 MHz originate from a region no larger than 204 km, {and the dipolar magnetic-field geometry model} indicates that the emission region centered at $\sim$800 km from the star. This is the tightest constraint on the size of the emission region reported so far for PSR B0329+54 at the given frequencies, and it is at least five times more stringent than the existing emission height constraints based on the dipolar geometry model.}

\section{Introduction}
\label{sec:intro}

Periodic signals from pulsars travel through the ionized plasma of the interstellar medium (ISM), causing a delay $\Delta\, t_{\nu}$ (in seconds) in a signal of frequency $\nu$ (in MHz) relative {to} an infinite frequency, which can be quantified using the cold plasma dispersion relation (valid for temperatures $\ll10^{4}$ K; e.g., \citealt{1985rpa..book.....R})

\begin{equation}
    \Delta t_{\nu} = 4.1488\times 10^{\,3}\,\,\text{DM} \,\,\nu^{-2} 
\label{DM_fit_equation}\end{equation}
($\nu^{-2}$ law from now on), where the dispersion measure (DM; in pc cm$^{-3}$) quantifies the line of sight (LOS) column free electron density.

The inhomogeneities in the ISM cause scatter-broadening of the radio signals from pulsars (e.g., \citealt{1971MNRAS.155...51S}); in addition, pulsars also exhibit profile evolution with frequency (e.g., \citealt{Layzer}). The scatter-broadening and the profile evolution with frequency become more evident {in large fractional bandwidth observations} at lower radio frequencies. Specifically, the exponential tail of the scattering can vary steeply as $\nu^{-4.4}$ for the Kolmogorov turbulence spectrum in the ISM (e.g., \citealt{1977ARA&A..15..479R}). This makes it challenging to track the reference or fiducial point in the pulsar profiles across different frequencies (e.g., \citealt{2007MNRAS.377..677A}). An ideal fiducial point in observed profiles represents the same rotational phase of the pulsar at all frequencies.

Intrinsic pulse evolution with frequency, scattering in the ISM, refraction within the pulsar magnetosphere (e.g., \citealt{2003A&A...412..473W}), and delays due to difference in heights of origin of different radio frequencies in the pulsar magnetosphere (e.g., \citealt{2001ApJ...555...31G}) can all lead to deviations from the $\nu^{-2}$ law in time-of-arrival (ToA) measurements. In the era of wide-band receiver telescopes, several investigations (e.g., \citealt{2022ApJ...930L..27K}, \citealt{2019A&A...624A..22D}, \citealt{2016ApJ...817...16C}, \citealt{2005MNRAS.357.1013A}, \citealt{1992ApJ...385..273P}) have shown that if a large frequency bandwidth is covered in the observations, the ToAs of radio pulses from pulsars may deviate from the $\nu^{-2}$ law. 

In this work, we focus on PSR B0329+54 (discovered by \citealt{1968Natur.219..574C}), which is one of the brightest ($\sim$1.5 Jy at 400 MHz; ATNF pulsar catalog\footnote{\url{https://www.atnf.csiro.au/research/pulsar/psrcat/}}; \citealt{2005AJ....129.1993M}) and the closest pulsars ($\sim$1.2 kpc; \citealt{2019ascl.soft08022Y}). It has been reported to show frequency-dependent DMs by \cite{2007MNRAS.377..677A}. Their findings show a DM difference of $\sim6\times10^{-3}$ pc cm$^{-3}$, {with DM precision $\sim 5\times 10^{-5}$ pc cm$^{-3}$,} between simultaneous observations of this pulsar at frequency combinations of 243$+$610 MHz and 320$+$610 MHz. 
    Another relevant study is \cite{2012A&A...543A..66H}, which reports a DM difference of $\sim2\times10^{-3}$ pc cm$^{-3}$ for PSR B0329+54 between the simultaneously observed frequency ranges of 42$-$78 MHz and 139$-$187 MHz, {however, the DM difference is not significant in this case given the error bars on DM (for 42-78 MHz) of $\pm10^{-3}$ pc cm$^{-3}$. In contrast to \cite{2007MNRAS.377..677A}, \cite{2012A&A...543A..66H} reported no clear signs of deviations from the $\nu^{-2}$ law.}

This study re-examines the case of PSR B0329$+$54 using the upgraded Giant Meterwave Radio Telescope\footnote{\url{http://www.gmrt.ncra.tifr.res.in}} (uGMRT; \citealt{2017JAI.....641011R}, \citealt{1997hsra.book..217S}). We observed this pulsar simultaneously across 300 to 1460 MHz, providing one of the most sensitive low-frequency datasets with broad frequency coverage for this pulsar to date. Using these observations, we aim to address the following three questions: (a) Do the delays caused by the ISM truly deviate from the $\nu^{-2}$ law? (b) Can a fiducial point in the pulsar profile be identified that consistently yields a single value of DM across all observed frequencies? (c) If such a point exists, how different {are its ToAs from the ToAs of} conventional pulsar timing methods, such as the cross-correlation methods that utilize full pulsar profiles (e.g., FFTFIT algorithm; \citealt{1992RSPTA.341..117T})?

\S 2 describes the observations and data reduction. \S 3 explains the various strategies used to determine ToAs. \S 4 presents the results, and \S 5 provides a comparison between the current findings and earlier models of PSR B0329+54. \S 6 addresses the study's conclusions.

\section{Observation and Data Reduction}
\label{sec:Section-2}

The pulsar observations were carried out using the uGMRT (as part of proposal DDTC346). The uGMRT comprises 30 parabolic dishes, each with a diameter of 45 meters, arranged in an approximately Y-shaped configuration. The uGMRT was selected for these observations due to its exceptionally wide bandwidth at lower radio frequencies (120-1460 MHz), where frequency-dependent effects in pulsar data are more prominent and can be studied in greater detail.

For the observation of PSR B0329+54, six of the central 1-square-kilometer antennas of the uGMRT were tuned to band-3 (300-500 MHz), eight other central antennas were set to band-4 (550-750 MHz), and the 16 arm antennas were configured to operate in band-5 (1260-1460 MHz).

The pulsar was observed simultaneously in band-3, band-4, and band-5, with a total integration time of $\sim$19 minutes. 
In each band, the Stokes-I phased array \texttt{filterbank} was recorded with a time resolution of 163.84 $\mu$s and 4096 frequency channels, providing a channel resolution of $\sim$48.83 kHz. Given the computational limitations of the uGMRT, the time resolution and frequency channel count are selected such that any millisecond-order deviations from the $\nu^{-2}$ law can be easily detected in the dataset while mitigating the effects of intrachannel smearing.

In all bands, the initial and final $\sim$5 MHz of data were mostly devoid of signal. In band-5, the first $\sim$25 MHz and the last $\sim$10 MHz were removed from the data due to strong radio frequency interference (RFI) contamination. Table \ref{Observational_set_up1} summarizes the observational setup described above.

\begin{table}[H]
    
    \centering
    \begin{adjustbox}{width=\columnwidth,center}
    \begin{tabular}{|c|c|c|c|c|c|c|c|}
    \hline
    
    uGMRT & Frequency & Usable& Time&  No. of &No. of\\
    Band &  range (MHz) & bandwidth (MHz) & Resolution ($\mu s$) & channels & Antennas\\
    \hline
    \hline
    Band-3 &  300-500 & $\sim$190 & 163.84 & 4096 & 6 \\
    Band-4 &  550-750 & $\sim$190 & 163.84 & 4096 & 8 \\
    Band-5 & 1260-1460 & $\sim$160 & 163.84 & 4096 & 16\\
    \hline
    \end{tabular}
    \end{adjustbox}
    \caption{The table lists the observational specifications for the simultaneous observation of PSR B0329+54 across the frequency range of 300-1460 MHz.}
    \label{Observational_set_up1}
\end{table}
PSR B0329$+$54 is a $\sim$714.5197 ms {(\citealt{2004MNRAS.353.1311H})} pulsar with a DM of $\sim$26.7641 pc cm$^{-3}$ {(\citealt{2012A&A...543A..66H})}, and its flux densities at 400 MHz, 600 MHz, and 1400 MHz are 1500 mJy, 1300 mJy, and 203 mJy {(\citealt{1995MNRAS.273..411L})}, respectively. For this pulsar, the maximum intra-channel smearing in the data is approximately $\sim$400 $\mu$s at the lowest frequency channel at 300 MHz. However, the smearing decreases to within the time resolution of the data for frequencies greater than $\sim$400 MHz. In addition, the maximum scattering timescale, estimated using equations (2) and (3) of \cite{1985ApJ...288..221C} for the Kolmogorov spectrum, is $\sim$5 $\mu$s at 300 MHz, which is negligible compared to the time resolution; thus, scattering effects are ignored in the subsequent analysis.

\begin{figure}[H]
\centering
        \includegraphics[width=0.5\linewidth,keepaspectratio]{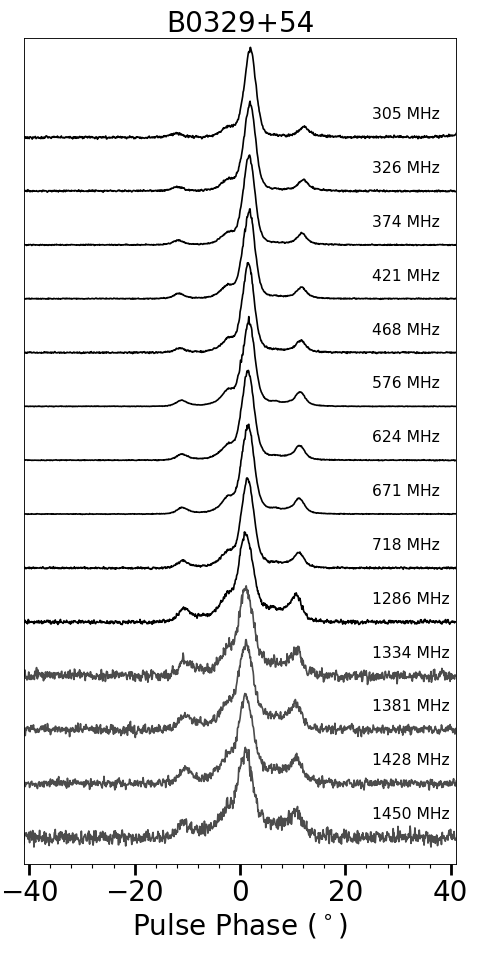}
    \caption{The figure showing the de-dispersed (at DM=26.74870(7) pc cm$^{-3}$) integrated profiles ($\sim$19 minutes) for PSR B0329+54, spanning frequencies from 300 to 1460 MHz. In band-3, 4, and 5 of the uGMRT, each with a 200 MHz bandwidth, the profiles are plotted for frequency channels located near the center of each consecutive 50 MHz subband. Each plotted profile corresponds to a single (out of 3$\times$4096) frequency channel with a bandwidth of $\sim$48.83 kHz. Additionally, the very first and last usable frequency channel profiles are included to demonstrate the broadest frequency range observation for this pulsar using the uGMRT.}
    \label{Dedisp_profiles}
\end{figure}

A total of $\sim$1594 pulses {($\sim$19 minutes)} were periodically folded for each frequency channel using the topocentric period of pulsar, $\sim$714.5515 ms, derived from \texttt{TEMPO2} {\citep{2006MNRAS.369..655H}}. The profile is divided into 4361 ($\sim$ pulsar period / time resolution) equally spaced bins. The de-dispersed (at DM = 26.74870(7) pc cm$^{-3}$; \S 4) integrated profiles for this pulsar, covering the frequency range of 300-1460 MHz, are presented in Figure \ref{Dedisp_profiles}.

\section{Methods for Determining {Times} of Arrival}
\label{sec:Section 3}
The PSR B0329$+$54 exhibits a three-component profile (Figure \ref{Dedisp_profiles}) across all observing frequencies. {\cite{2011ApJ...741...48C} investigated mode changing phenomenon in this pulsar at 610 and 1540 MHz, where the leading component can sometimes become very faint at low frequencies, and the trailing component can sometimes be challenging to distinguish from the central peak at high frequencies. In our observations, however, all three components remain clearly detectable and distinguishable from 300-1460 MHz, suggesting a mode where these effects may not have an impact.} 

The components are defined as follows: the first (smallest, leading), the second (strongest, central), and the third (in between in size, trailing), referred to as Comp 1, Comp 2, and Comp 3, respectively, as shown in Figure \ref{Template_profile}. The key aspect of this analysis is determining the ToAs of fiducial points in pulsar profiles across all observed frequencies from 300 to 1460 MHz. The following set of points is considered as fiducial markers:

\subsection{Peaks in the Profile and the Midpoint of the Outer Peaks}
\vspace{-2mm}
\subsubsection{Peak of the Strongest Component}
\vspace{-2mm}

The peak of the profile is the most easily identifiable fiducial point in pulsar profiles. \cite{1969ApL.....3..225R} is one of the earliest studies to utilize the profile peak as a fiducial marker. In this analysis, the ToAs of the profile peak have been determined by fitting a 3-point parabola at the location of the peak, similar to \cite{1992ApJ...385..273P}, of Comp 2.

\vspace{-8mm}

\subsubsection{Individual Peaks of the Leading and Trailing Components}
\vspace{-2mm}

In addition to tracing the strongest peak, the ToAs are also determined for the peaks of Comp 1 and Comp 3. Again, the ToAs for these components are found by fitting a 3-point parabola at the location of each peak.

\vspace{-8mm}

\subsubsection{Midpoint of the Peaks of the Leading and Trailing Components}
\vspace{-2mm}

The midpoint of the two peaks in the profile is a commonly used fiducial point in pulsars with two-component profiles (as first suggested by \citealt{1970PhDT.........8C}). To determine the ToAs, the midpoint between the peaks of Comp 1 and Comp 3 is used as the fiducial point.

\subsection{Times of Arrival of the FFTFIT Algorithm; Template Method}
The FFTFIT Algorithm \citep{1992RSPTA.341..117T} is a well-established method for determining ToAs in high-precision pulsar timing studies (e.g., pulsar timing array experiments searching for nanohertz gravitational waves; \citealt{2023ApJ...951L...8A}). This algorithm relies on a template profile that closely resembles the pulsar's true integrated profile, assuming that the observed profile is a scaled and shifted version of this template. The ToAs are determined using a chi-square minimization approach, comparing the Fourier-transformed template with the observed pulsar profile. For a detailed explanation of the technique, refer to Appendix A of \cite{1992RSPTA.341..117T}.

\begin{figure}[H]
\centering
        \includegraphics[width=0.5\linewidth,keepaspectratio]{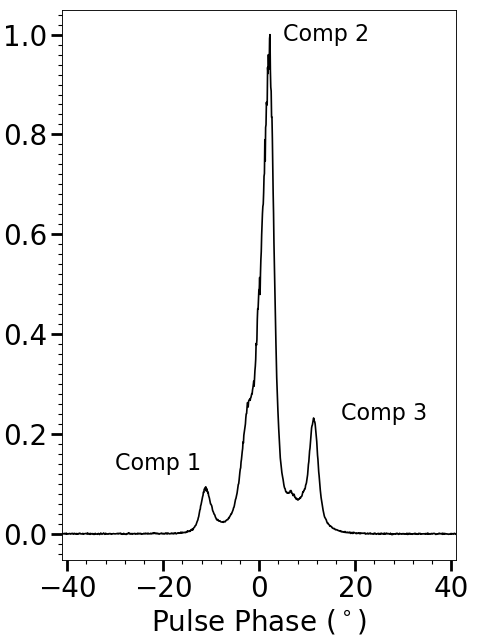}
    \caption{Figure showing the normalized template profile for PSR B0329$+$54 utilized for extracting ToAs in the FFTFIT algorithm.}
    \label{Template_profile}
\end{figure}
In this analysis, the highest S/N ($\sim$1555 for the peak of Comp 2) profile, corresponding to a single channel with a frequency of $\sim$589 MHz, was used as the template (Figure \ref{Template_profile}). The PAT command of the PSRCHIVE software \citep{2012AR&T....9..237V} is used to determine FFTFIT-based ToAs from the observed pulsar profiles. We have used a single profile template to determine ToAs across profiles of all frequency channels from 300 to 1460 MHz. Note that a more robust approach involves using frequency-dependent templates to extract the ToAs from individual frequency bands, which can help mitigate the effects of profile evolution with frequency in the extracted ToAs. The methods described in {\citealt{2019A&A...624A..22D}, \citealt{2020A&A...644A.153D}, and \citealt{2022PASA...39...53T}} provide a comprehensive description of the use of frequency-dependent templates for high-precision pulsar timing experiments. Since this method utilizes the {template} profile for ToA determination, it will be referred to as `template' in the following context.

\section{Results}

\subsection{Fitting ToAs with a $\nu^{-2}$ model}

\begin{figure}[H]

\centering
        \includegraphics[width=1.0\linewidth,keepaspectratio]{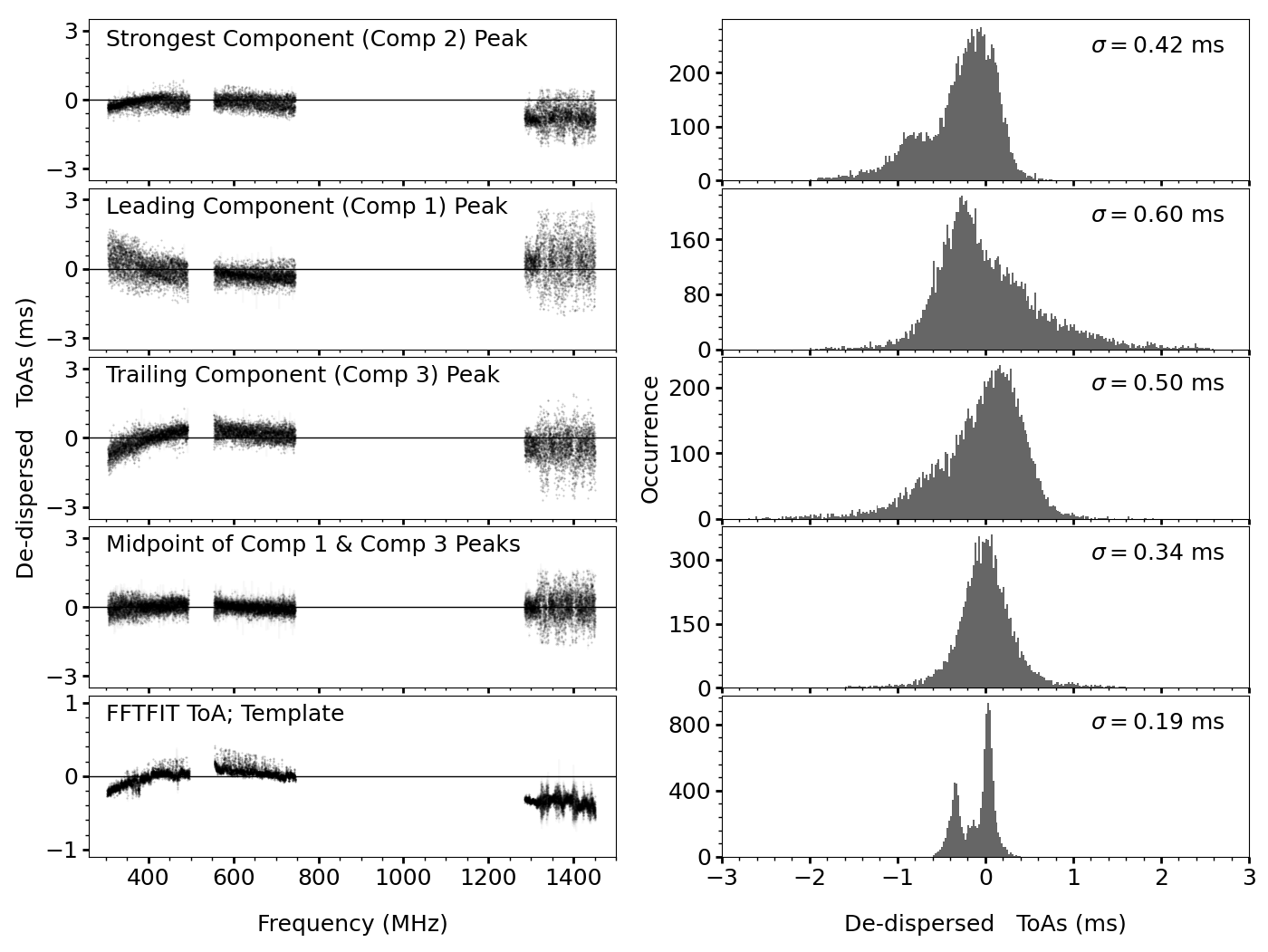}
    \caption{Left: Figure showing the ToAs after removing the dispersive delays for different frequencies, using the best-fitted DM, and a constant time offset. Each plot indicates the name of method and its corresponding best-fitted DM value. Right: Histogram showing the spread of de-dispersed ToAs (from the left plots) for different methods.}
    \label{Residual_ToAs_different_DMs}
\end{figure}

The ToAs for all the described methods were extracted for Band-3, Band-4, and Band-5, and Table \ref{Number_of_ToAs} lists the number of ToAs and the median ToA error for each band; the ToAs and the observing frequencies are Doppler-corrected for Earth's motion around the Sun. For each method, a single DM (using equation \ref{DM_fit_equation}) and a constant time offset are fitted simultaneously in the ToAs from 300–1460 MHz. Figure \ref{Residual_ToAs_different_DMs} shows the residuals of the ToAs (observed ToAs$-$fitted model) for individual methods. The best-fit DM values for individual methods are listed in Table \ref{Number_of_ToAs}.

\begin{table}[H]
    \centering
    \begin{adjustbox}{width=\columnwidth,center}
    \begin{tabular}{|c|c|c|c|}
    \hline
    Methods & Number of ToAs & Median ToA Error ($\mu$s)& Best Fitted DM \\
    & (Band-3, Band-4, Band-5) & (Band-3, Band-4, Band-5) & (pc cm$^{-3}$) \\
    \hline
    \hline
    Comp 2 Peak & 3921, 3861, 3320 & 15, 11, 42 & 26.78076(1) \\
    \hline
    Comp 1 Peak & 3750, 3846, 3106 & 46, 45, 45 & 26.67192(3) \\
    \hline
    Comp 3 Peak & 3822, 3900, 3298 & 43, 38, 46 & 26.82530(3) \\
    \hline
    Mid Point of & 3759, 3893, 3151 & 100, 91, 99 & 26.74870(7) \\
    Comp 1 \& Comp 3 & & & \\
    \hline
    Template & 3876, 3803, 3331 & 15, 11, 30 & 26.78077(1) \\
    \hline
    \end{tabular}
    \end{adjustbox}
    \caption{Table presents the median ToA error, the number of usable ToAs per band for various methods, and the best-fit DM values. ToAs with large error bars ($\sim$ few ms) or deviating significantly from the fitted $\nu^{-2}$ curve were excluded, causing $\sim$2$-$6$\%$ fluctuations in the number of ToAs per band for different methods.}
    \label{Number_of_ToAs}
\end{table}

The spread (standard deviation) of residuals, as represented by histograms in Figure \ref{Residual_ToAs_different_DMs}, for Comp 1, Comp 2, and Comp 3 are 0.60 ms, 0.42 ms, and 0.50 ms, respectively. Given the relative strength of each component peak, as observed from Figure \ref{Template_profile}, {it shows that} a higher S/N of the fiducial marker {is} associated with a smaller residual spread. Additionally, using the midpoint of Comp 1 and Comp 3 results in a reduced spread of 0.34 ms, which is less than that of the individual components. Furthermore, employing the full profile for the template method results in the least spread of 0.19 ms in the residuals.

Applying a single DM across 300$-$1460 MHz results in systematics (deviations from normal distribution) in the residuals of Comp 1, Comp 2, Comp 3, and the template method (Figure \ref{Residual_ToAs_different_DMs}). Remarkably, the residuals corresponding to the midpoint of Comp 1 and Comp 3 align under a single DM across the full frequency range, with the residual histogram closely approximating a normal distribution. {In the case of the template method, no single DM is valid for all frequency bands simultaneously. However, given the spread of ToAs for the current observations, it remains a possibility that the residuals for the template method might fall within those of the midpoint of the peaks of Comp 1 and Comp 3. However, from this point onwards, we will only focus on whether the single-valued DM for the midpoint has any unique features.}


\subsection{De-dispersing All ToAs with a Single DM}

The absolute ToAs for individual fiducial markers are separated by timescales of the order of ms, providing a means to differentiate between various residuals. For this purpose, all the ToAs for different fiducial markers, including the ToAs of template method, were de-dispersed using a single DM, corresponding to the midpoint of Comp 1 and Comp 3 (Table \ref{Number_of_ToAs}), without applying any constant time offset. Figure 4 shows the absolute ToAs for individual methods after de-dispersion.

\begin{figure}[H]

\centering
        \includegraphics[width=1.0\linewidth,keepaspectratio]{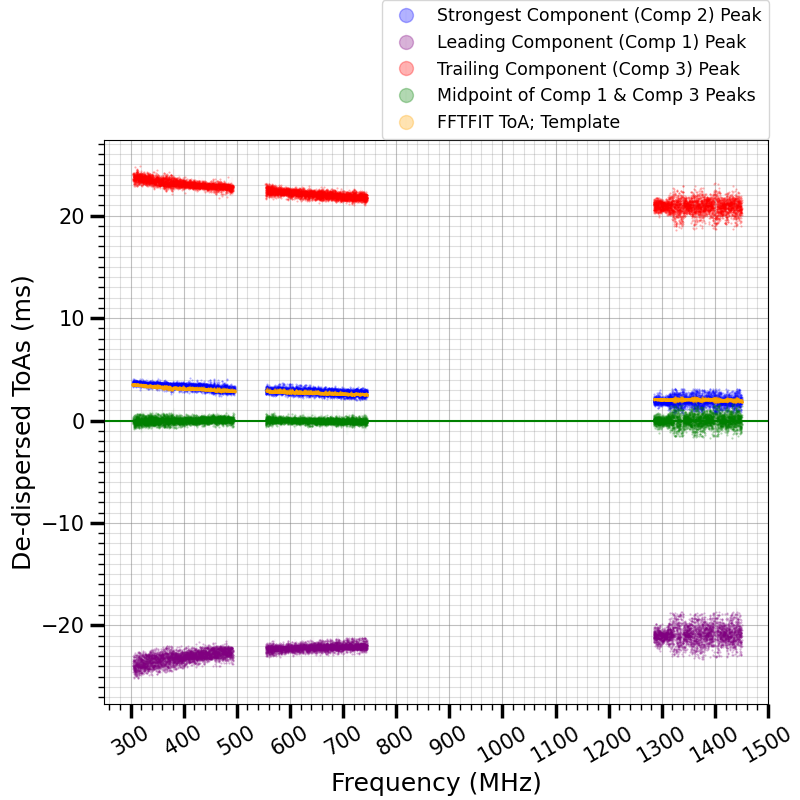}
    \caption{Figure showing the de-dispersed (at DM = 26.74870(7) pc cm$^{-3}$) ToAs for a set of fiducial markers along with the ToAs of template method. The zero line corresponds to MJD of $\sim$60426.494074723545.}
    \label{Figure4}
\end{figure}

In Figure \ref{Figure4}, the ToAs corresponding to the mid-points of comp 1 and comp 3 (in green) are positioned along the middle horizontal line. Notably,  the ToAs of the template method (in orange) align with those of Comp 2 (in blue). {The FFTFIT method utilizes the full profile to extract the ToAs, with the ToAs being effectively determined by the steepest features in the profile (\citealt{2006ApJ...642.1004V}). The observed alignment between ToAs of FFTFIT and Comp 2 may be related to the fact that component 2 contains the sharpest and longest steep features in the profile, and it also dominates the full profile.} Additionally, the ToA error bars for both the template method and the Comp 2 peak are also very close to each other (Table \ref{Number_of_ToAs}). The ToAs corresponding to the peaks of Comp 1 (in purple) and Comp 3 (in red) {appear to }exhibit a symmetry around the reference line {for PSR B0329+54}. The key observation from this figure is the presence of a point (specifically, the midpoint of the Comp 1 and Comp 3 peaks) in the profile of PSR B0329+54 where all frequency signals from 300 to 1460 MHz appear to {arrive approximately} at the same time and show a {good approximation to the} $\nu^{-2}$ law.

\subsection{Investigating Causes of Frequency-Dependent DM}

From Figure \ref{Residual_ToAs_different_DMs}, it is clear that using a fiducial marker other than the mid-point of Comp 1 and Comp 3 can result in frequency-dependent DMs {for PSR B0329+54}. 

Another potential {consequence} of frequency-dependent DMs is the spread (of the order of hundreds of $\mu$s) in the de-dispersed ToAs. While the DM for the mid-point of Comp 1 and Comp 3 appears unique, it is essential to determine what the minimum observation bandwidth should be to achieve a DM value that is valid across all frequency ranges.

\begin{figure}[H]

\centering
        \includegraphics[width=1.0\linewidth,keepaspectratio]{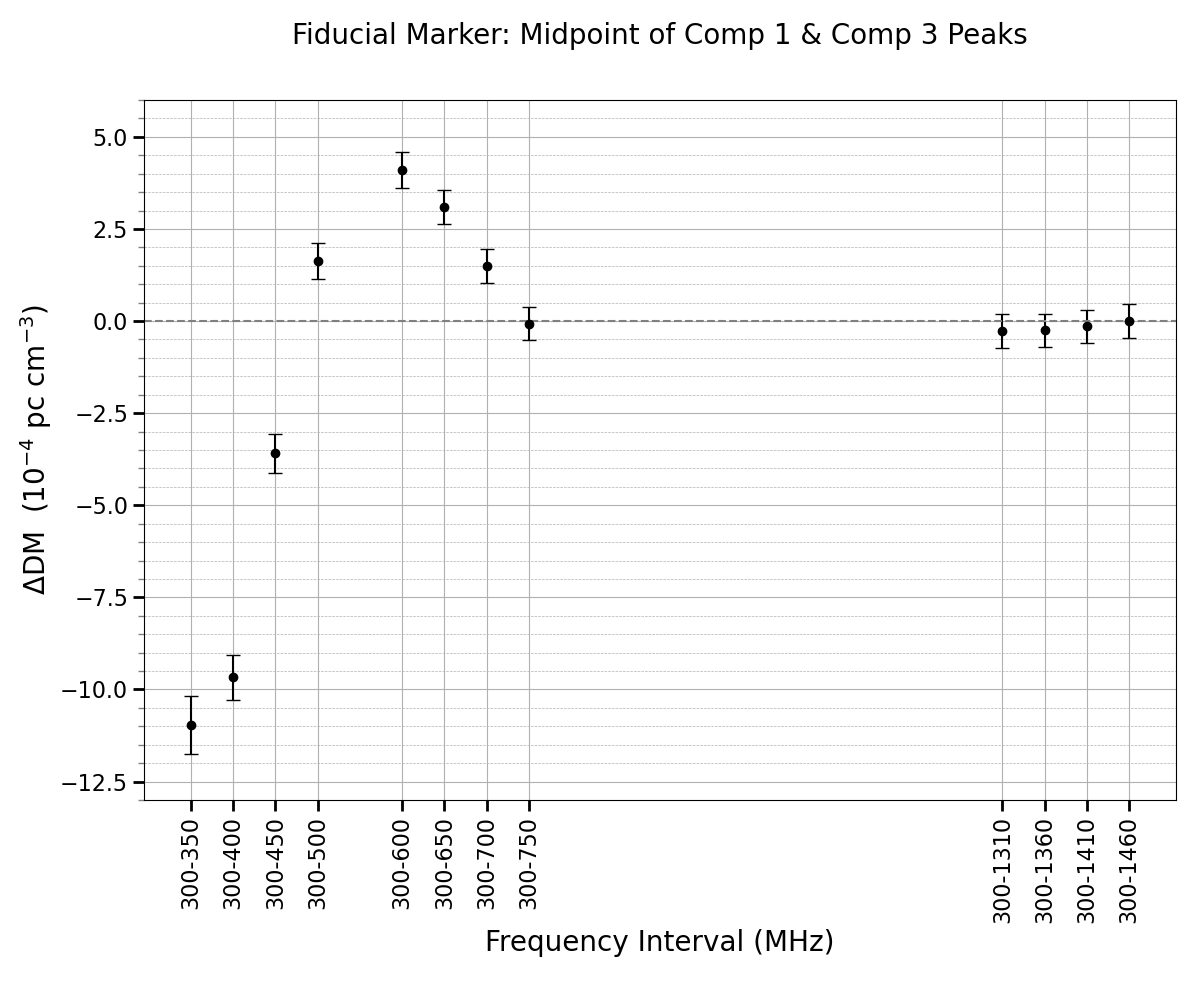}
    \caption{The figure shows the fitted DM values for subsets of ToAs (for the mid-point of Comp 1 and Comp 3) corresponding to progressively increasing frequency intervals. The zero line represents a DM of 26.74870 pc cm$^{-3}$.}
    \label{Freq-dependent_DM}
\end{figure}

Figure \ref{Freq-dependent_DM} shows the fitted values of DM for subsets of ToAs (for the mid-point of Comp 1 and Comp 3), corresponding to progressively increasing observational frequency bandwidths. After subtracting a reference MJD ($\sim$60426.494074723545) from the absolute ToAs, the DM was fitted for each subset of ToAs using equation \ref{DM_fit_equation}.

Figure \ref{Freq-dependent_DM} indicates that a unique value of DM is achieved when the observing frequency bandwidth exceeds $\sim$400 MHz (starting from 300 MHz). {For the mid-point of peaks of Comp 1 and Comp 3}, using a narrower frequency range, {starting from 300 MHz}, can lead to a DM discrepancy of the order of $10^{-3}$ pc cm$^{-3}$ between different frequencies {for PSR B0329+54. Additionally, the obtained DM values include the effects of profile evolution with frequency. }

These findings highlight the need for observations across a wide frequency range to effectively investigate the phenomenon of frequency-dependent DMs. {Another observation from Figure 5 is that the DM is not changing after including the Band-5 (1260-1460 MHz) ToAs into the DM fitting, nor is there any improvement in the DM precision. This feature may be related to the limited sensitivity of the uGMRT Band-5 observations included in this work. It suggests the need for further investigation of the DM convergence observed in Figure 5 using telescopes with higher sensitivity near 1.4 GHz.}

\section{Revisiting Emission Models of PSR B0329+54}

\begin{figure}[H]

\centering
        \includegraphics[width=0.8\linewidth,keepaspectratio]{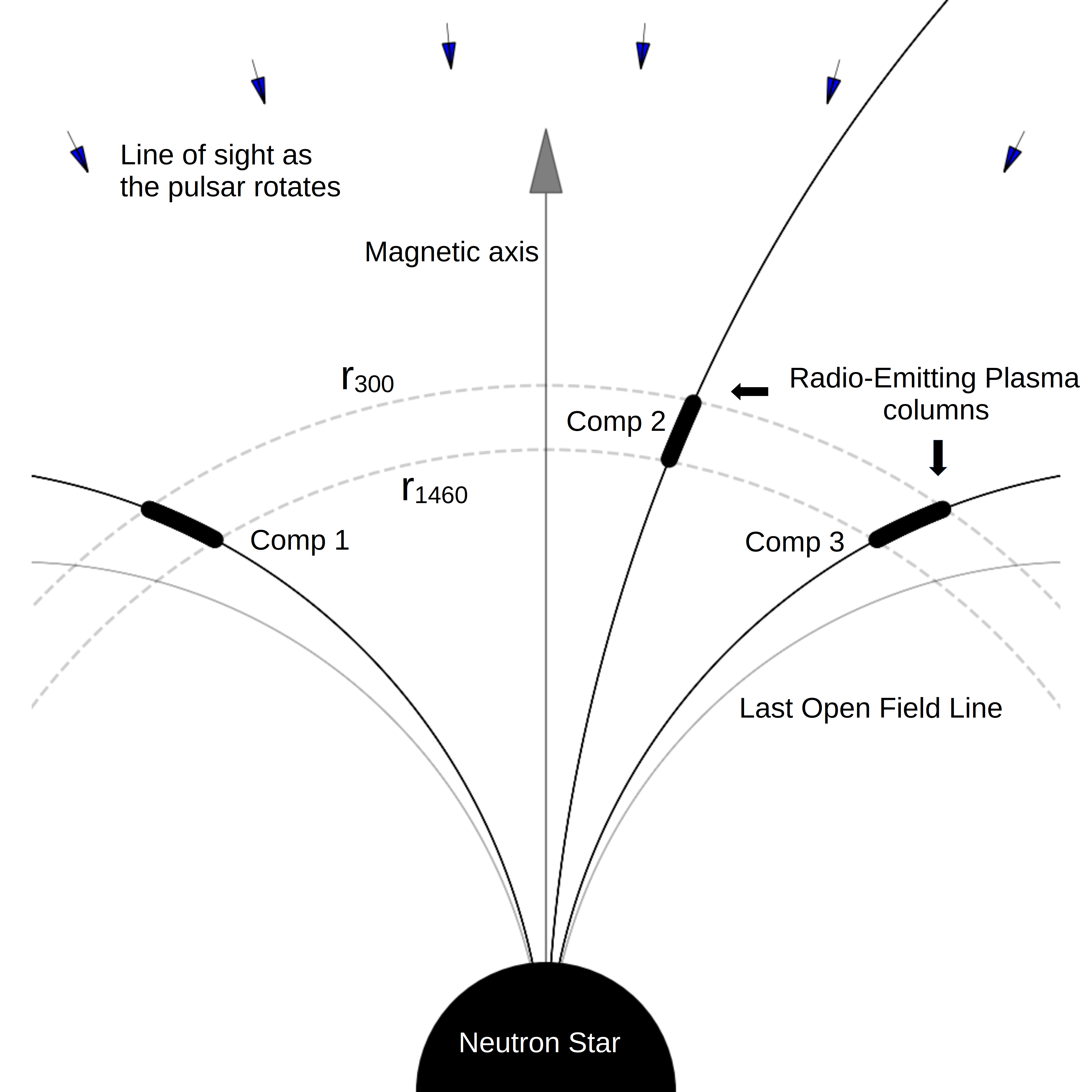}
    \caption{Figure illustrating a schematic diagram of the PSR B0329+54. Here, $r_{300}$ and $r_{1460}$ correspond to the emission heights at frequencies of 300 MHz and 1460 MHz, respectively.}
    \label{Schematic_diagram}
\end{figure}

\cite{1986A&A...163..114K} conducted a single pulse study of PSR B0329+54 at 102.5 and 1700 MHz, finding a linear relationship between the ToAs at these two frequencies. This correlation was observed to be consistent across the ToAs of all components of this pulsar. Based on these findings, \cite{1991A&A...243..219G} proposed a geometrical model to explain the correlation between the ToAs at two distinct frequencies. The model proposed in \cite{1991A&A...243..219G} is as follows: The radio emission phenomenon is coherent curvature radiation, in which the emission originates from charged particle bunches as they move along dipolar magnetic field lines, with each radio frequency emitted from a narrow region \citep{1975ApJ...196...51R}. Higher frequencies are emitted from low altitudes near the star, while lower frequencies originate from relatively higher altitudes (e.g., \citealt{1991ApJ...370..643B}). Figure \ref{Schematic_diagram} presents a schematic diagram of the radio emission region corresponding to three components of PSR B0329+54 at frequencies from 300 to 1460 MHz.

The relation between ToAs at two frequencies, $\nu_l$ and $\nu_h$ ($\nu_l < \nu_h$), is described by

\begin{equation}
    t_{ih}=t_{il}\left(\frac{r_{h}}{r_{l}}\right)^{1/2}+\left(\frac{1+sin\,\theta}{c}\right)[r_{l}-r_{h}]
\label{ToA_equation_2}\end{equation}
(\cite{1991A&A...243..219G}), where, $t_{il}$ and $t_{ih}$ represent the ToAs (in seconds) of the i-th component of pulsar profile at frequencies $\nu_l$ and $\nu_h$, respectively, which originate at heights $r_l$ and $r_h$ (in km, with $r_l > r_h$) from the center of the star. The angle $\theta$ denotes the separation (in radians) between the magnetic and rotational axes, and c is the speed of light ($\sim 3\times10^{5}$ km/s). Applying this model to the observations reported in \cite{1986A&A...163..114K}, \cite{1991A&A...243..219G} found that the emission height for a given radio frequency remains constant across different pulse components. Considering this result along with equation \ref{ToA_equation_2}, the average ratio of heights for two frequencies can be determined using multiple components by \begin{equation}
\left<\frac{r_{h}}{r_{l}}\right>=\frac{1}{2N}\sum_{\substack{i=1\\ i \neq j}}^N \sum_{\substack{j=1}}^N\left(\frac{t_{ih}-t_{jh}}{t_{il}-t_{jl}}\right)^2
\label{ratio_equation}\end{equation}
where N represents the number of components in the profile. Equation \ref{ToA_equation_2} can also be modified to find the average height difference using multiple components by:
\begin{equation}
   \left<r_{l}-r_{h}\right>= \frac{1}{N}\left(\frac{c}{1+sin\,\theta}\right)\sum_{\substack{i=1}}^N\left|t_{ih}-t_{il}\left<\frac{r_{h}}{r_{l}}\right>^{1/2}\right|
\label{diff_equation}\end{equation}

The frequency-dependent effects induced by the ISM are supposed to be uniform across all components of the pulsar. Consequently, the different systematic variations observed in the ToAs for the three components, as shown in Figure \ref{Figure4}, must arise from the intrinsic geometry of the emission-dominated magnetic field lines. The ToAs (relative to the midpoint of Comp 1 and Comp 3) for Comp 1, 2, and 3 for the lowest ($\sim$305 MHz) and highest ($\sim$1450 MHz) observed frequencies are [$-23.629\pm 0.042$, $3.883 \pm 0.028$, $24.003\pm 0.035$] ms, and [$-21.407\pm 0.045$, $1.618\pm 0.046$, $20.712\pm 0.042$] ms, respectively.

If the midpoint of Comp 1 and Comp 3 coincide with the magnetic axis of PSR B0329+54, we can use equation \ref{ratio_equation} and ToAs for the three components at 305 and 1450 MHz to obtain the height ratio $r_{1450}/r_{305}=0.794\pm 0.003$. Substituting this value into equation \ref{diff_equation} along with $\theta\sim30^{\circ}$ (\citealt{1990ApJ...352..247R}, \citealt{1988MNRAS.234..477L}) result in a height difference $r_{305}-r_{1450}= 191 \pm 12 $ km. Using the height ratio and the height difference, the individual heights can be determined as $r_{305}=931\pm 61$ km and $r_{1450}= 739\pm 60$ km. Note that the error bars on the height ratio, height difference, and absolute heights represent only the fitting errors propagated from the ToA errors in Equations \ref{ratio_equation} and \ref{diff_equation}. These equations assume that all components have the same heights at a given frequency. However, as observed in Figure \ref{Residual_ToAs_different_DMs}, the ToAs seem to arrive simultaneously for all frequencies at the mid-point of peaks of Comp 1 and Comp 3. Given this, a more physically meaningful upper bound on the height difference can be determined by considering the spread of ToAs at the mid-point of peaks of Comp 1 and Comp 3. The spread, $\sigma = \pm0.34$ ms (Figure \ref{Residual_ToAs_different_DMs}), translates to a distance of $|2\,\sigma\,c| \sim 204$ km, thus $|r_{305} - r_{1450}| \leq 204$ km, with the absolute heights potentially ranging somewhere between $r_{1450}^{min} = 679$ km and $r_{305}^{max} = 992$ km. 

\begin{figure}[H]

\centering
        \includegraphics[width=0.9\linewidth,keepaspectratio]{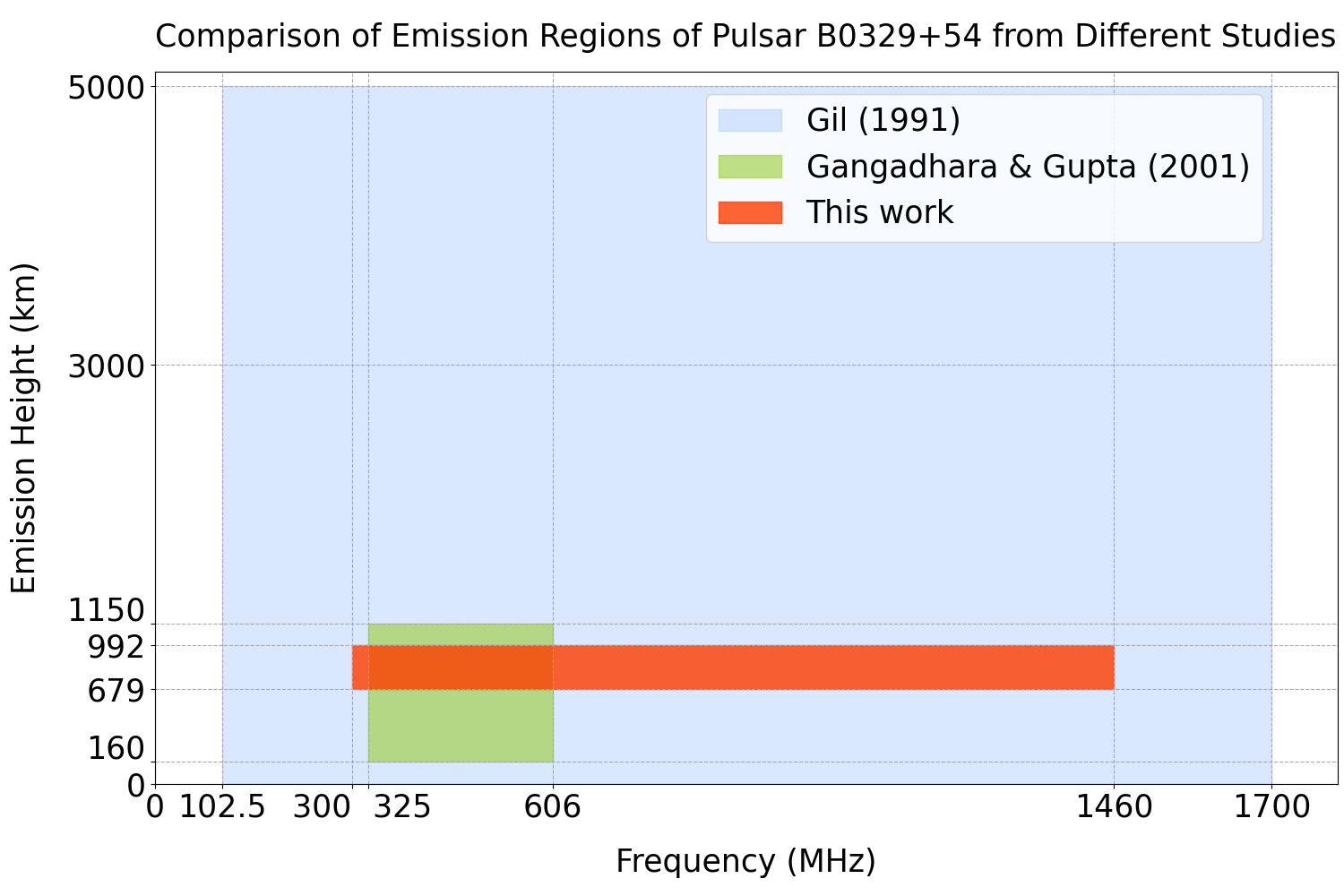}
    \caption{The figure shows a comparison of emission heights across different frequency ranges from earlier studies (see text) and this work.}
    \label{Emission_heights}
\end{figure}

\cite{1991A&A...243..219G} reported the upper limits on emission heights for PSR B0329+54 at frequencies 102.5 and 1700 MHz, specifically $r_{102}\leq5000$ km and $r_{1700}\leq3000$ km, respectively. While providing the upper limits, \cite{1991A&A...243..219G} suggests that the actual emission heights can be as close as 300 km from the star, with a height difference of $\sim$200 km between the two frequencies. Moreover, \cite{2001ApJ...555...31G} reported emission heights derived from single-pulse studies of PSR B0329+54 at frequencies of 325 MHz and 606 MHz, with an inferred height range of $\sim$160 to 1150 km. The estimated emission heights from both studies are consistent with the values obtained in this work, though this study places a more stringent constraint on the emission heights, as shown in Figure \ref{Emission_heights}.

The errors in DM values used for the dedispersion of the profile are covariant with the derived emission heights. \cite{1991A&A...243..219G} reported that the DM used for dedispersion is less precise than $\sim 10^{-2}$ pc cm$^{-3}$, while for \cite{2001ApJ...555...31G}, Figure \ref{Freq-dependent_DM} illustrates that their DM can be inaccurate by $\sim 10^{-3}$ pc cm$^{-3}$ due to their narrow frequency range observations. The potential reason for the improved constraints on emission heights in the current study is the enhanced frequency sampling between 300 and 1460 MHz, resulting in a DM value with a precision of $\sim 10^{-5}$ pc cm$^{-3}$ (for the mid-point of peaks of Comp 1 and Comp 3). {This DM measurement assumes that the mid-point of the peaks of Comp 1 and Comp 3 corresponds to a unique and stable phase for all observed frequencies, which physically corresponds to a location near the magnetic axis of PSR B0329+54. This assumption is based on the approximately zero delay behavior observed between different frequencies for the de-dispersed ToAs corresponding to the mid-points of the peaks of Comp 1 and Comp 3, as shown in Figure \ref{Figure4}.}




In Figure \ref{Figure4}, the average value of $|t_{305}-t_{1450}|\sim2.6 $ ms for the ToAs of three components. Also, the spread of ToAs of the mid-point of peaks of Comp 1 and Comp 3, $\pm 0.34$ ms, provides an upper limit on the delay due to height differences between 305 and 1450 MHz. If the height difference between the two frequencies remains consistent across all three components, the primary contribution to $|t_{305}-t_{1450}|$  may be attributed to the delay arises as the observer's line of sight crosses different emission regions in the pulsar magnetosphere as the pulsar rotates.


Further single-pulse studies, such as those by \cite{1986A&A...163..114K}, \cite{2001ApJ...555...31G}, and \cite{Mitra_2024}, utilizing more sensitive, high-bandwidth observations, may reveal pulse correlations across different radio frequencies, offering deeper insights into potential emission height differences.

\section{Conclusions}

The findings of this study are as follows:

1) For PSR B0329+54, the ToAs for the peaks of Comp 1, 2, 3, and the FFTFIT method exhibit ms$-$order deviations from the $\nu^{-2}$ law for the frequency range of 300$-$1460 MHz. {The ToAs derived from the} FFTFIT method align with those of the highest S/N component in the profile.

2) {We find that the midpoints of the peaks for Comp 1 and Comp 3 align most closely with the $\nu^{-2}$ law, in comparison to the other approaches employed in this work. Note that we have used a single profile template in the FFTFIT method to extract ToAs across the 300–1460 MHz frequency range.}

3) The fitted DM for the midpoint of peaks of Comp 1 and Comp 3 ($26.74870(7)$ pc cm$^{-3}$) significantly differ ($\sim 4\times10^{-2}$ pc cm$^{-3}$) from the DM value obtained using the FFTFIT method ($26.78077(1)$ pc cm$^{-3}$).

4) The ToA {spread} is dependent on the S/N of the fiducial marker; a higher S/N corresponds to a lower {spread} of the peak of an individual component. The use of the full profile or multiple components results in less {spread} compared to that of the individual peaks.

5) {We find that for the mid-point of the peaks of Comp 1 and Comp 3, the DM reaches a stable value once the observation bandwidth exceeds 400 MHz, starting from 300 MHz, for PSR B0329+54. In contrast, using a narrower bandwidth can result in discrepancies in DM measurements of the order of 10$^{-3}$ pc/cm$^{-3}$.}

6) The de-dispersed ToAs for the peaks of Comp 1 and Comp 3 exhibit symmetry around the de-dispersed ToAs for their midpoint. This may be visualized as the midpoint of peaks of Comp 1 and Comp 3 aligned with the magnetic axis of PSR B0329+54, with Comp 1 and Comp 3 originating from two symmetrical dipolar magnetic field lines around the magnetic axis, while Comp 2 originates from a field line close to the magnetic axis.

7) This work establishes a tight upper bound of $\sim$204 km on the emission height difference between 300 and 1460 MHz, which is at least $\sim5$ times better than the earlier investigations. Assuming that all components of pulsar originate from the same height for a given frequency and utilizing the dipolar geometry model, the absolute emission heights for 300-1460 MHz are estimated to lie between $\sim$ 680 and 990 km.

8) This study suggests that the observed millisecond order deviation from $\nu^{-2}$ law may not arise due to differences in emission heights for various frequencies, but rather that the primary contribution comes as the observer's line of sight crosses different emission regions in the pulsar magnetosphere as the pulsar rotates.

{One of the key findings of this study is that the widely used FFTFIT timing technique can consistently produce ToAs that align closely with the sharpest or steepest feature in the profile, even when using a single profile template across a broad frequency range (300–1460 MHz) as utilized in this work. Furthermore, the incorporation of frequency-dependent templates {should} enhance the robustness of the ToAs because {they should no longer be} affected by profile evolution with frequency.} This work provides insights into the potential causes {of measurements} of the frequency-dependent DMs and encourages further investigation in pulsars where this effect has been observed. Additionally, it addresses the significance of exploring alternative fiducial markers to effectively remove the delays due to ISM and enhance the precision in pulsar timing experiments.

\section{Acknowledgments}

We would like to express our gratitude to \textbf{Prof. Dipanjan Mitra} (NCRA-TIFR) for highlighting the problem of radius-to-frequency mapping in PSR B0329+54 to the authors of this article and for providing guidance from time to time during the progress of this work. We also extend our thanks to Mr. Sanjay Kudale and GMRT operator Mr. Deepak for their assistance during the observations with the GMRT. Additionally, we appreciate Dr. Tim Pennucci for sharing his valuable insights on timing techniques. We thank Dr. Utkarsh Kumar for useful discussion during the progress of this work. SSS acknowledges the support of the National Science and Technology Council of Taiwan through grant number 113-2123-M-001-008-.
TH acknowledges the support of the National Science and Technology Council of Taiwan through grants 113-2112-M-005-009-MY3 and 113-2123-M-001-008-. SY acknowledges the support from NSTC through grant numbers 113-2112-M-005-007-MY3 and 113-2811-M-005-006-. SH is supported by the Australian Research Council (ARC) Centre of Excellence (CoE) for Gravitational Wave Discovery (OzGrav) project numbers CE170100004 and
CE230100016, and the ARC CoE for All Sky Astrophysics in 3 Dimensions (ASTRO 3D) project number CE170100013.

{We sincerely thank the anonymous reviewer for their detailed and thoughtful review of our manuscript.}

\end{document}